\documentclass[twocolumn]{webofc}

\usepackage[varg]{txfonts}   
\usepackage{graphicx}
\begin{document}

\title{Commissioning of miniBELEN-10A, a moderated neutron counter with a flat efficiency for thick-target neutron yields measurements}
\author{\firstname{N} \lastname{Mont-Geli}\inst{1}\fnsep\thanks{\email{nil.mont@upc.edu}} \and
        \firstname{A} \lastname{Tarifeño-Saldivia}\inst{2}\fnsep\thanks{\email{atarisal@ific.uv.es}} \and
        \firstname{L M} \lastname{Fraile}\inst{3} \and
        \firstname{S} \lastname{Viñals}\inst{4} \and
        \firstname{A} \lastname{Perea}\inst{5} \and
        \firstname{M} \lastname{Pallàs}\inst{1} \and
        \firstname{G} \lastname{Cortés}\inst{1} \and
        \firstname{G} \lastname{Garcia}\inst{4} \and
        \firstname{E} \lastname{Nácher}\inst{2} \and
        \firstname{J L} \lastname{Tain}\inst{2} \and
        \firstname{V} \lastname{Alcayne}\inst{6} \and
        \firstname{O} \lastname{Alonso-Sañudo}\inst{3} \and
        \firstname{A} \lastname{Algora}\inst{2} \and
        \firstname{J} \lastname{Balibrea-Correa}\inst{2} \and
        \firstname{J} \lastname{Benito}\inst{3} \and
        \firstname{M J G} \lastname{Borge}\inst{5} \and
        \firstname{J A} \lastname{Briz}\inst{5} \and
        \firstname{F} \lastname{Calviño}\inst{1} \and
        \firstname{D} \lastname{Cano-Ott}\inst{6} \and
        \firstname{A} \lastname{De Blas}\inst{1} \and
        \firstname{C} \lastname{Domingo-Pardo}\inst{2} \and
        \firstname{B} \lastname{Fernández}\inst{7,8} \and
        \firstname{R} \lastname{Garcia}\inst{1} \and
        \firstname{J} \lastname{Gómez-Camacho}\inst{7,8} \and
        \firstname{E M} \lastname{González-Romero}\inst{6} \and
        \firstname{C} \lastname{Guerrero}\inst{7,8}
        \firstname{J} \lastname{Lerendegui-Marco}\inst{2} \and
        \firstname{M} \lastname{Llanos}\inst{3} \and
        \firstname{T} \lastname{Martínez}\inst{6} \and
        \firstname{V} \lastname{Martínez-Nouvilas}\inst{3} \and
        \firstname{E} \lastname{Mendoza}\inst{6} \and
        \firstname{J R} \lastname{Murias}\inst{3} \and
        \firstname{S E A} \lastname{Orrigo}\inst{2} \and
        \firstname{A} \lastname{Pérez de Rada}\inst{6} \and
        \firstname{V} \lastname{Pesudo}\inst{6} \and
        \firstname{J} \lastname{Plaza}\inst{6} \and
        \firstname{J M} \lastname{Quesada}\inst{7} \and
        \firstname{A} \lastname{Sánchez}\inst{6} \and
        \firstname{V} \lastname{Sánchez-Tembleque}\inst{3} \and
        \firstname{R} \lastname{Santorelli}\inst{6} \and
        \firstname{O} \lastname{Tengblad}\inst{6} \and
        \firstname{J M} \lastname{Udías}\inst{3} \and
        \firstname{D} \lastname{Villamarín}\inst{6}
}
\institute{Institut de Tècniques Energètiques (INTE), Universitat Politècnica de Catalunya (UPC), E-08028, Barcelona, Spain 
\and Instituto de Física Corpuscular (IFIC), CSIC – Univ. Valencia (UV), E-46071, Valencia, Spain 
\and Grupo de Física Nuclear (GFN) and IPARCOS, Universidad Complutense de Madrid (UCM), E-28040, Madrid, Spain
\and Centro de Micro-Análisis de Materiales (CMAM), Universidad Autónoma de Madrid (UAM), E-28049, Madrid, Spain
\and Instituto de Estructura de la Materia (IEM), CSIC, E-28006 Madrid, Spain
\and Centro de Investigaciones Energéticas, Medioambientales y Tecnológicas (CIEMAT), E-28040, Madrid, Spain
\and Departamento de Física Atómica, Molecular y Nuclear, Universidad de Sevilla (US), E-41012 Sevilla, Spain
\and Centro Nacional de Aceleradores (CNA), Universidad de Sevilla (US) - J. Andalucía - CSIC, E-41092, Sevilla, Spain
}

\abstract{miniBELEN-10A is a modular and transportable moderated neutron counter with a nearly flat detection efficiency up to 8 MeV. The detector was designed to carry out measurements of ($\alpha,n$) reactions in the context of the Measurement of Alpha Neutron Yields (MANY) project. In this work we present the results of the commissioning of miniBELEN-10A using the relatively well-known thick-target neutron yields from $^{27}$Al($\alpha,n$)$^{30}$P.}

\maketitle

\section{Introduction}
\label{intro}
The production of neutrons through $\alpha$-induced reactions plays a crucial role in many different fields. Specifically, in nuclear astrophysics, ($\alpha,n$) reactions are the major source of neutrons for the slow neutron-capture nucleosynthesis process (the s-process) \cite{RevModPhys.83.157} and are involved in the weak rapid neutron-capture process (named the weak r-process, also known as the $\alpha$-process) \cite{Bliss_2017}. Moreover, ($\alpha,n$) reactions are important in underground physics, as they are one of the main sources of the neutron-induced background in underground facilities, which is typically a crucial issue for the experiments carried out there \cite{westerdale2022alpha}. Other fields of interest include nuclear technologies such as fission \cite{BEDENKO2019189} and fusion reactors \cite{Cerjan_2018} and non-destructive assays for non-proliferation and spent fuel management applications \cite{broughton2021sensitivity,romano2020alpha}.

All these applications rely on accurate and precise experimental data. However, most of the available data on ($\alpha,n$) reactions was measured decades ago, is incomplete or presents large discrepancies not compatible with the declared uncertainties. Furthermore, in many cases the uncertainties on the cross-sections and neutron energy spectra are rather large. To address such problems new measurements are required \cite{westerdale2022alpha}. To that end the Measurement of Alpha Neutron Yields (MANY) collaboration was formed. The project relies on the employment of the currently existing infrastructure in Spain, in particular the $\alpha$-particle beams produced by the accelerators at CMAM (Madrid) \cite{redondo2021current} and CNA (Sevilla) \cite{gomez2021research}; and on the use of neutron detection systems such as miniBELEN, which is a moderated neutron counter based on the use of $^3$He-filled detectors, and MONSTER, which is a time-of-flight neutron spectrometer based on the use of BC501/EJ301 liquid scintillation modules \cite{garcia2012monster}. Both systems are complemented by $\gamma$-spectroscopy measurements using an array of fast LaBr$_3$(Ce) scintillation detectors of the FATIMA type \cite{vedia2017performance} with angular resolution capabilities and a high-purity germanium detector.

\section{miniBELEN}
\label{miniBELEN}
The miniBELEN detector builds upon the experience gained from previous experiments involving other detectors of the BELEN-type \cite{gomez2011first,agramunt2014new,tarifeno2017conceptual}. It is a 4$\pi$ moderated neutron counter which is based on the use of a set of $^3$He-filled thermal neutrons proportional counters embedded in a modular block of High-Density PolyEthylene (HDPE) which acts as moderator. Modularity means that the HDPE moderator is formed by smaller blocks which can be reassembled in different ways in order to provide the detector with different types of response. In practice, in the case of miniBELEN, this means that we have three detectors or configurations for the prize of one. These configurations have been designed in order to present a neutron detection efficiency nearly independent from the neutron energy, namely a flat efficiency, up to 8 MeV with a slightly negative slope up to 10 MeV. 

\begin{figure}[h]
\centering
\includegraphics[width=5 cm,angle=0]{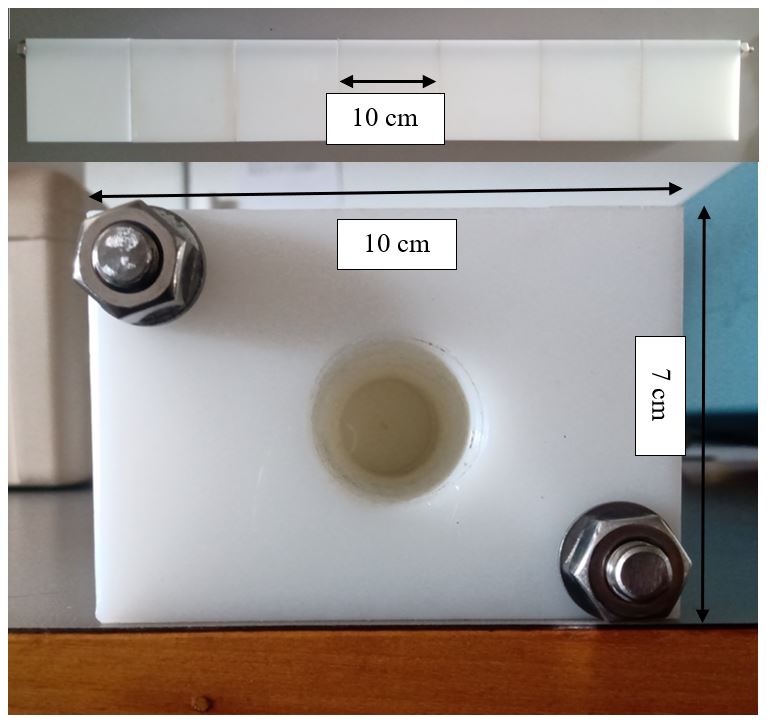}
\caption{HDPE block from miniBELEN (7x10x70 cm$^3$) with central hole to be filled with a neutron counter. Each of these blocks is made of seven smaller blocks (7x10x10 cm$^3$) which are assembled using two stain-less steel rods.}
\label{figSingleBlock}     
\end{figure}

The basic components of miniBELEN are the 7x10x70 cm$^3$ HDPE blocks with a central hole where the neutron counters (60 cm length and 1 inch diameter) are embedded. Each of these blocks is made of seven smaller 7x10x10 cm$^3$ blocks which are assembled using two stainless-steel rods as it is shown in figure \ref{figSingleBlock}.

\begin{figure}[h]
\centering
\includegraphics[width=7 cm]{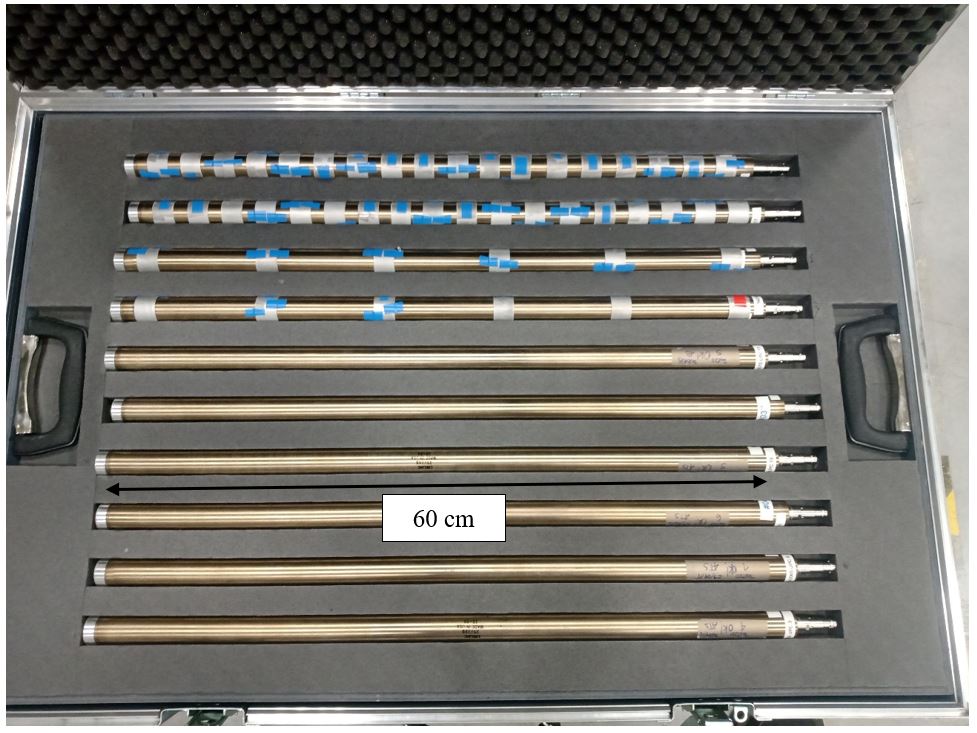}
\caption{The ten $^3$He-filled neutron proportional counters of miniBELEN-10A (60 cm length and 1 inch diameter). Some of the counters are partially covered with cadmium filters (2 cm length and 0.5 mm thickness) in order to provide the detector with a flat neutron efficiency.}
\label{cadmium}     
\end{figure}

In miniBELEN the flat efficiency is achieved by partially covering some of the neutron counters with cadmium filters in order to optimize the contribution of each ring (i.e., a group of counters which are all placed at the same distance from the centre of the detector) to the total detection efficiency (see figure \ref{cadmium}). This is what we call the composition method.

\begin{figure}[ht]
\centering
\includegraphics[width=6 cm]{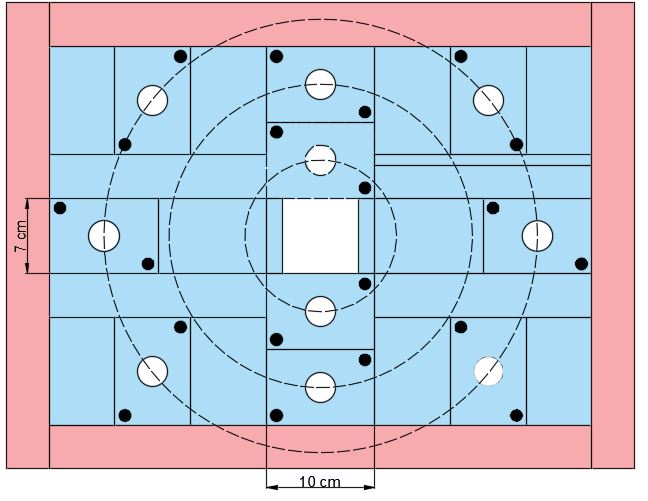}
\caption{Schematic plot of the HDPE moderator of miniBELEN-10A. The total volume is 58x43x70 cm$^3$. The length of the red blocks is 50 cm while the length of the blue blocks is 70 cm. The area of the central hole (white) is 7x7 cm$^2$. The white circles represent the position of the proportional counters and the black circles represent the stainless-steel rods. Dashed circles: rings of counters.}
\label{figHDPEblocks}     
\end{figure}

In his work we present the results from the commissioning of miniBELEN-10A, which is a configuration of the miniBELEN detector using ten cylindrical $^3$He-filled proportional counters manufactured by LND \cite{lnd}. The gas pressure is generally 10 atm although in three counters of the outer ring the gas pressure is  4, 8 and 20 atm. Figure \ref{figHDPEblocks} shows a schematic plot of the modular HDPE moderator used in miniBELEN-10A. 

\section{Detector efficiency}
\label{deteff}
\begin{figure}[h]
\centering
\includegraphics[width=7.5 cm]{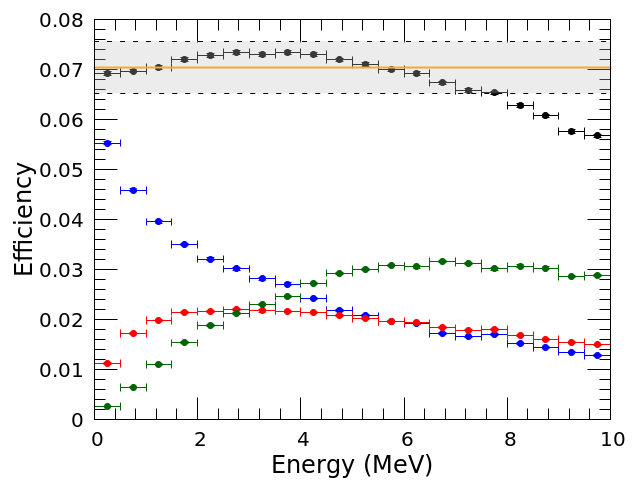}
\caption{Monte Carlo calculated neutron detection efficiency of miniBELEN-10A. The y-axis error-bars are smaller than the size of the data points. Black: total efficiency. Red: efficiency of ring 1. Blue: ring 2. Green: ring3. Orange line: nominal efficiency calculated according to equation \ref{nomEff}. Grey band (limited by the black dashed lines): nominal error-band calculated according to equation \ref{nomErr}.}
\label{figeff}     
\end{figure}

Monte Carlo calculations were carried out using \textit{Particle Counter} \cite{particlecounter}, which is a Geant4 application \cite{agostinelli2003geant4}, in order to determine the neutron detection efficiency of miniBELEN-10A. These calculations were carried out using a point-like and isotropic neutron source placed at the center of the detector. Results are shown in figure \ref{figeff} in the form of a 0.5 MeV bin-size histogram.

The nominal efficiency of miniBELEN-10A ($\varepsilon$) is defined as the average efficiency within a certain energy range (from 0 up to $E_{Max}$),

\begin{equation}
  \label{nomEff}
  \varepsilon = \frac{1}{M}\sum_{k = 0}^{E_{Max}}\varepsilon_k = 0.0705
\end{equation}

Being $\varepsilon_k$ the detection efficiency of bin $k$ and $M$ the total number of bins. In the case of miniBELEN-10A the nominal efficiency is calculated up to $E_{Max}$ = 8 MeV. It is worth noting that the alpha-energies involved in this experiment were not larger than 9.3 MeV and that the $Q$-value of $^{27}$Al($\alpha,n$)$^{30}$P is 2.64241(8) MeV. Consequently, the maximum energy reached by the emitted neutrons will always be lower than 8 MeV. 

In order to take into account the efficiency fluctuations, the uncertainty of the nominal efficiency ($\delta\varepsilon$) is defined in the following way,

\begin{equation}
  \label{nomErr}
  \delta\varepsilon = \max(\varepsilon_k^{max}-\varepsilon,\varepsilon-\varepsilon_k^{min}) = 0.0052
\end{equation}

Being $\varepsilon_k^{max}$ and $\varepsilon_k^{min}$, respectively, the maximum and minimum values of the neutron detection efficiency between 0 and $E_{Max}$. 

An experimental characterization of miniBELEN-10A was carried out using several $^{252}$Cf neutron sources. Due to the lack of a well-calibrated neutron source the detection efficiency was determined using the Neutron Multiplicity Counting (NMC) technique \cite{Henzlova_2019,langner1998application}. Relative discrepancies between the experimental efficiencies and Monte Carlo simulations using a point-like and isotropic neutron source were not larger than 5\%. In these simulations the $^{252}$Cf neutron emission energy spectrum from reference \cite{radev2014neutron} was used.

\section{Commissioning of miniBELEN-10A}
\label{comissioning}
The commissioning was carried out by measuring the relatively well-known $^{27}$Al($\alpha,n$)$^{30}$P thick-target neutron yields at Centro de Micro-Análisis de Materiales (CMAM, Madrid, Spain). The alpha-particles beam (charge +2e) was accelerated using the 5 MV tandem accelerator at CMAM and was then passed through a stainless steel cylindrical collimator which also acted as secondary electrons suppression system after applying a -100 V bias voltage. The collimator was covered with a natural tantalum ($^{181}$Ta) thick foil in order to avoid the production of neutrons. It must be taken into account that the ($\alpha,n$) threshold for natural tantalum is 10.009(5) MeV. 

The target consisted on a high-purity (99.9995\%) natural aluminium ($^{27}$Al) thick foil which was placed at the center of the detector immediately after the collimator. The beam-current was measured by collecting the charge reaching the target and integrating it with an ORTEC-439 digital current integrator. The data acquisition system (DACQ) is based on the use of a SIS3316 sample digitizer (16 channels and 250 MHz sampling frequency) from Struck Innovative Systeme. The digitizer is controlled by the \textit{Gasific 7.0} software \cite{AGRAMUNT201669}.
\begin{figure}[h]
\centering
\includegraphics[width=8 cm]{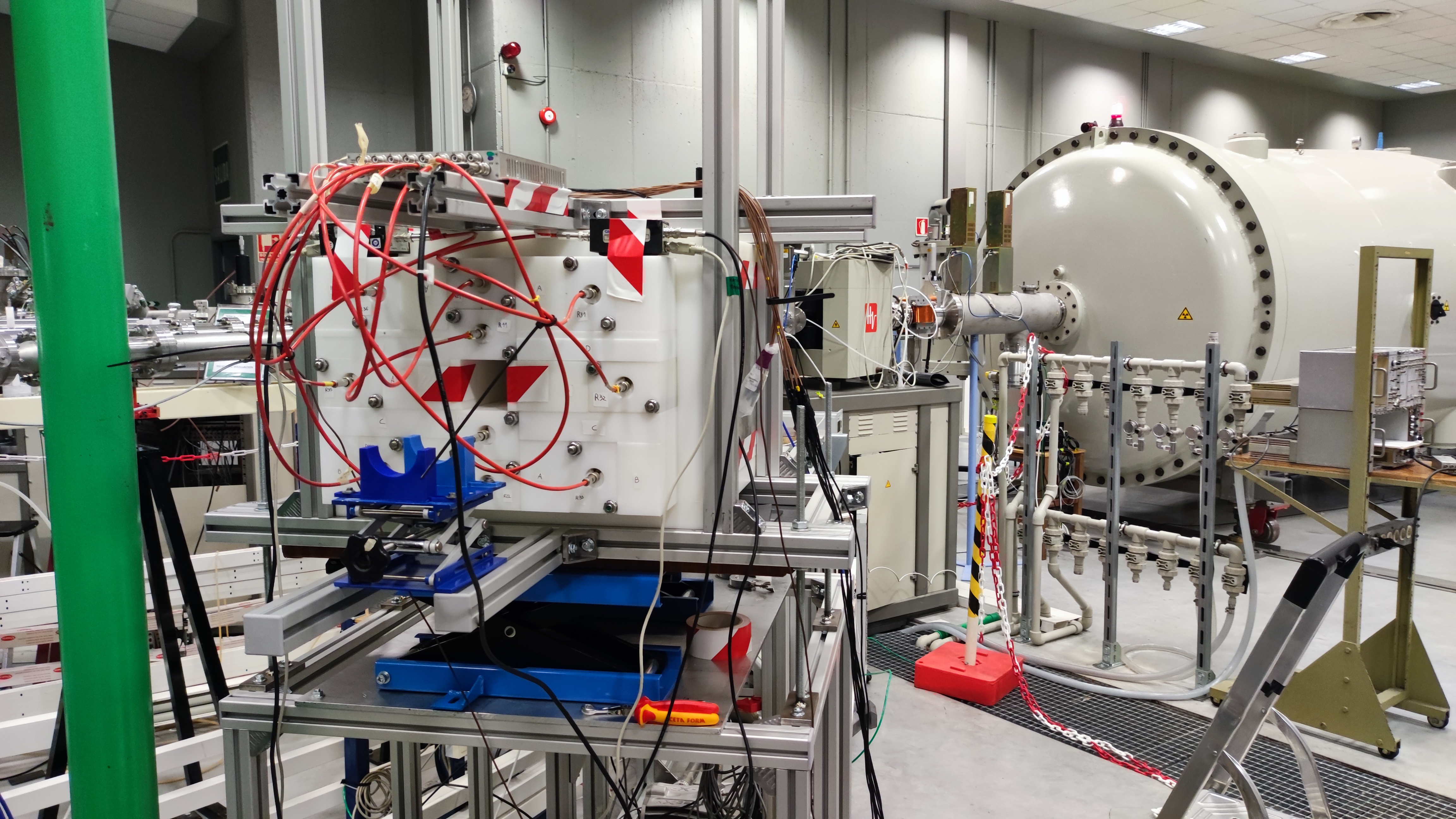}
\caption{Experimental setup at CMAM with miniBELEN-10A on the foreground and the 5MV tandem accelerator behind it.}
\label{figSetup}     
\end{figure}

\subsection{Thick-target yields}
\label{yields}
At a given beam-energy $E_\alpha$, the thick-target neutron production yields $Y(E_\alpha)$ can be determined in the following way,

\begin{equation}
  \label{eq:yields}
  Y(E_\alpha) = \frac{\sum_{i = 1}^{10} r_i}{\varepsilon I}
\end{equation}

Being $r_i$ the neutron counting rate in counter $i$, $\varepsilon$ the nominal neutron detection efficiency and $I$ the beam-current expressed in particles per time unit. 

The neutron background was estimated by carrying out measurements with a dummy target (natural tantalum) at 3.6, 3.7, 5.0, 5.5, 6.0, 7.4 and 8.5 MeV. A cubic spline interpolation of the data points was used in order to get the background yields at each energy relevant for the experiment. For measurements at beam-energies larger than 5 MeV the background contribution was found to be negligible (lower than 1\%). For energies between 3.6 and 3.7 MeV the contribution of the background yields ranged from 16\% up to 30\%. 
\begin{figure}[h]
\centering
\includegraphics[width=8.5 cm]{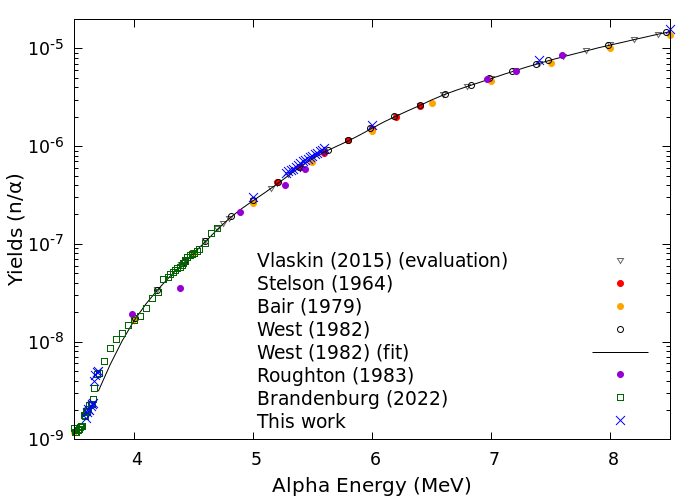}
\caption{$^{27}$Al($\alpha,n$)$^{30}$P thick-target neutron yields. Blue-crosses: measurement with miniBELEN-10A. Circles: data from references \cite{stelson1964cross,bair1979neutron,west1982measurements,ROUGHTON1983341,brandenburg2022measurements}. Grey triangles: evaluation from reference \cite{vlaskin2015neutron}. Black line: cubic spline fit to data from reference \cite{west1982measurements}.}
\label{Alyields}     
\end{figure}

When the production yields are large enough, dead-time effects become relevant and the counting rates in miniBELEN-10A, unless corrected, are underestimated. Our experience suggests that for counting rates below 10 kilo-counts per second and channel, the behavior of the DACQ can be modeled as a nonparalyzable system so that the correction model for such type of acquisition systems can be applied. 

The thick-target neutron yields from $^{27}$Al($\alpha,n$)$^{30}$P measured with miniBELEN-10A are shown in figure \ref{Alyields}. The relative errors are about 7\% and are due to the uncertainty of the nominal efficiency. The plot also includes data from previous measurements carried out by other groups (via direct neutron counting \cite{stelson1964cross,bair1979neutron,west1982measurements,brandenburg2022measurements} or activation \cite{ROUGHTON1983341}) and the evaluation from Vlaskin \textit{et al.} \cite{vlaskin2015neutron}. Our measurements are consistent with most of the previously existing experimental data within the margin of uncertainty. 

\section{Summary and final remarks}
\label{summary}
miniBELEN is a modular moderated neutron counter based on the use of $^3$He-filled detectors with a nearly flat response up to 8 MeV with a slightly negative slope up to 10 MeV. Modularity means that the moderator can be reassembled in different ways in order to obtain different types of response. In this work we have presented the results of the commissioning of configuration miniBELEN-10A through the measurement of the relatively well-known thick-target neutron yields from $^{27}$Al($\alpha,n$)$^{30}$P. Assuming a flat nominal efficiency determined through Monte Carlo calculations, we have been able to reproduce the results from previous experiments carried out by other groups.\newline

\small
This work has been supported by the Spanish Ministerio de Economía y Competitividad under grants FPA2017-83946-C2-1 and C2-2, PID2019-104714GB-C21 and C22, PID2021-126998OB-I00 and RTI2018-098868-B-I00, the Generalitat Valenciana Grant PROMETEO/2019/007, both cofounded by FEDER (EU), and the SANDA project funded under H2020-EURATOM-1.1 Grant No. 847552. The authors acknowledge the support from Centro de Micro-Análisis de Materiales (CMAM) - Universidad Autónoma de Madrid, for the beam time proposal (\textit{Commissioning of neutron detection systems for ($\alpha,n$) reaction measurements}) with code P01156, and its technical staff for their contribution to the operation.

\bibliography{ProceedingND22_v1}

\end{document}